\documentclass[pra,reprint,showpacs,raggedbottom,nofootinbib,superscriptaddress,aps,floatfix]{revtex4-1}
\usepackage{bm,dcolumn}
\usepackage{amsmath,amsfonts,amssymb,amsthm,amscd,braket} 
\usepackage{graphicx}
\usepackage{verbatim}
\usepackage{lipsum}
\usepackage[section]{placeins}
\usepackage{color}
\usepackage{xcolor}
\usepackage{soul}

\begin{document}
\sloppy
\title{Electromagnetically induced transparency in inhomogeneously broadened solid media}
\author{H. Q. Fan}
\affiliation{United States Army Research Laboratory, Adelphi, Maryland 20783, USA}
\affiliation{Joint Quantum Institute, University of Maryland, College Park, Maryland 20742, USA}
\author{K. H. Kagalwala}
\affiliation{Joint Quantum Institute, University of Maryland, College Park, Maryland 20742, USA}
\author{S. V. Polyakov}
\affiliation{National Institute of Standards and Technology, Gaithersburg, Maryland 20899, USA}
\author{A. L. Migdall}
\affiliation{Joint Quantum Institute, University of Maryland, College Park, Maryland 20742, USA}
\affiliation{National Institute of Standards and Technology, Gaithersburg, Maryland 20899, USA}
\author{E. A. Goldschmidt}
\email{elizabeth.a.goldschmidt2.civ@mail.mil}
\affiliation{United States Army Research Laboratory, Adelphi, Maryland 20783, USA}
\begin{abstract}
We study, theoretically and experimentally, electromagnetically induced transparency (EIT) in two different solid-state systems. Unlike many implementations in homogeneously broadened media, these systems exhibit inhomogeneous broadening of their optical and spin transitions typical of solid-state materials. We observe EIT lineshapes typical of atomic gases, including a crossover into the regime of Autler-Townes splitting, but with the substitution of the inhomogeneous widths for the homogeneous values. We obtain quantitative agreement between experiment and theory for the width of the transparency feature over a range of optical powers and inhomogeneous linewidths. We discuss regimes over which analytical and numerical treatments capture the behavior. As solid-state systems become increasingly important for scalable and integratable quantum optical and photonic devices, it is vital to understand the effects of the inhomogeneous broadening that is ubiquitous in these systems. The treatment presented here can be applied to a variety of systems, as exemplified by the common scaling of experimental results from two different systems. 
\end{abstract}
\maketitle
\section{Introduction}

Coherent processes in atomic ensembles are the basis for many implementations of quantum memory, coherent control of atomic populations, and mediation of interactions between optical fields \cite{MF05}. Electromagnetically induced transparency (EIT) is a canonical example of such a process with applications including slow and stopped light \cite{turukhin2001observation, schraft2016stopped,longdell2005stopped}, atomic-based field sensing \cite {fleischhauer2000quantum, katsoprinakis2006high}, lasing without inversion \cite{wu2008evidence,JGB14}. Most studies of EIT and other coherent processes in atomic ensembles have been conducted in gaseous media  over a range of temperatures from ultracold quantum gases to heated vapor cells \cite{MF05,IN12}, while a relatively smaller effort has been made in solid-state \cite{turukhin2001observation,BSH97,schraft2016stopped,RA17}. In fact, it was originally thought EIT would be impossible in solids \cite{SEH97}. However, atom-like emitters in solids offer benefits for quantum optical processes including higher densities, freedom from motional dephasing, and the possibility of integrated photonics approaches \cite{zhong2017nanophotonic}. The density of emitters in a solid can be as large as $\mathrm{10^{22}~cm^{-3}}$ while retaining atom-like optical properties \cite{ahlefeldt2016ultranarrow}. Solid-state ensembles of rare-earth atoms, in particular, are a promising platform for quantum memory and other applications due to their long spin coherence times \cite{zhong2015optically}. 

A major difference between solid-state systems and atomic gas systems is the static inhomogeneity of both the optical and spin transitions common in solid-state ensembles due to variations in the local electromagnetic field environment at each emitter location. Some of this variation is from strain due to material defects and imperfections, but even with high-purity materials, the inhomogeneous linewidth of the optical transition for an ensemble of solid-state emitters is often orders of magnitude larger than the homogeneous linewidth of each emitter. This inhomogeneity has potential benefits, particularly for the possibility of spectral multiplexing \cite{JN08, PhysRevA.86.053837}, but it also complicates coherent processes like EIT. To date, most studies of the effect of inhomogeneity on EIT and other quantum optical processes have focused on Doppler broadened gases \cite{AJ02,CYY02,EF06}. But inhomogeneously broadened solids present a different situation where motional effects are not present and the coherence time is not limited by transit time broadening \cite{EK02}.

\begin{figure}[ht]
  \centering
  \includegraphics[width=\columnwidth]{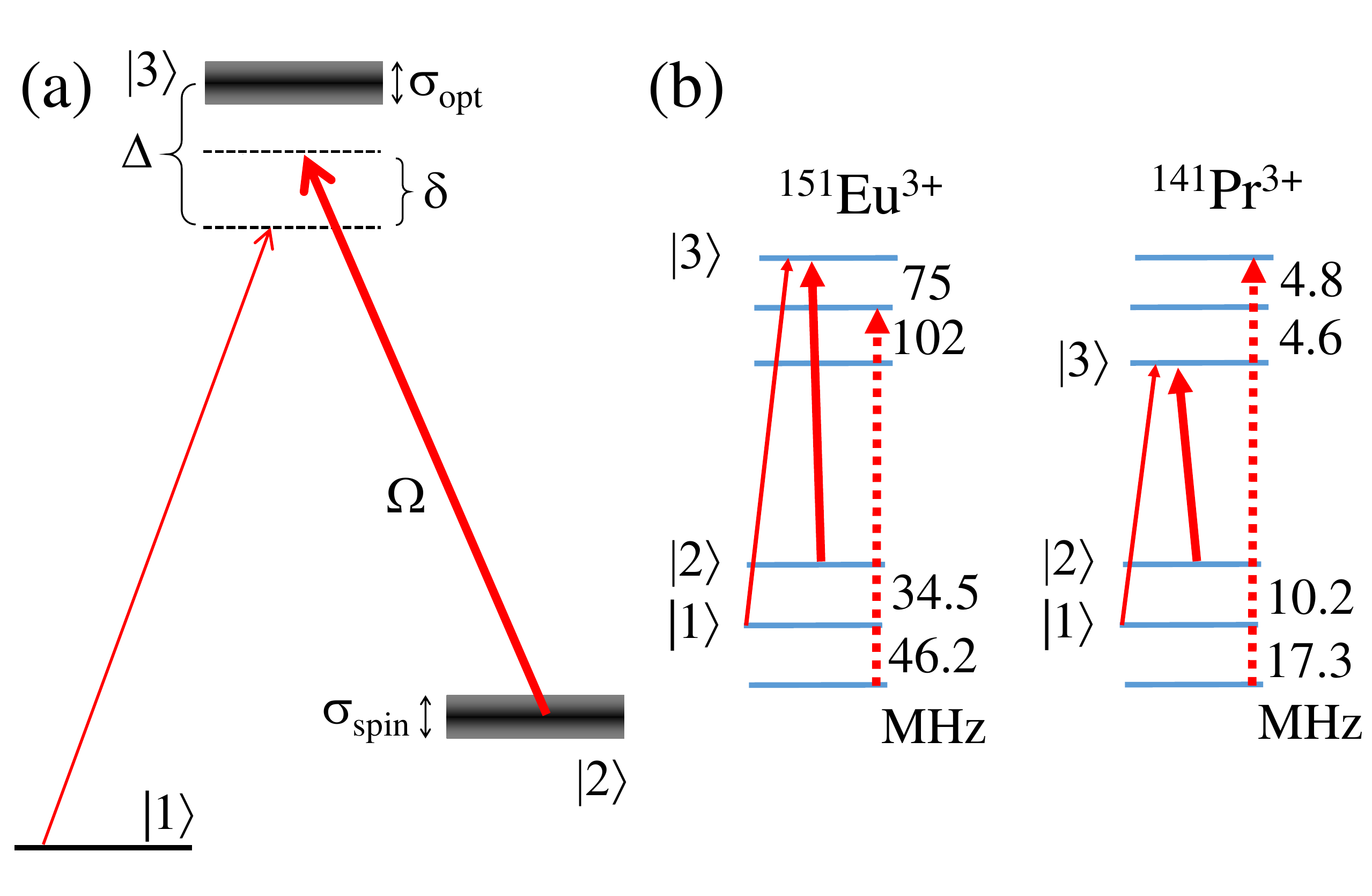}
\caption{(a) Energy level diagram for a $\Lambda$-system. The weak probe field (thin arrow) is detuned from the $|1\rangle\rightarrow|3\rangle$ transition by $\Delta$. The strong coupling field (thick arrow) on the $|2\rangle\rightarrow|3\rangle$ transition has Rabi frequency $\Omega$ and the difference between the probe and coupling detunings, the two-photon detuning, is $\delta$. 
The optical and spin transitions have inhomogeneously broadened widths  $\sigma_{\mathrm{opt}}$ and $\sigma_{\mathrm{spin}}$, respectively. (b) Energy level diagrams for Eu:YSO and Pr:YSO. The three states that make up the $\Lambda$-system in each case are labeled, as is the transition (dashed arrow) used for repumping during spectral hole-burning.}
  \label{fig:levels}
\end{figure}

We report results of $\mathrm{\Lambda}$-type EIT in two rare-earth doped solids, yttrium orthosilicate doped with europium (Eu:YSO) and with praseodymium (Pr:YSO) \cite{GL06}. We use spectral hole-burning techniques to control the optical inhomogeneous linewidth \cite{GP00,MN04}. We note that while the homogeneous optical linewidth is much larger in Pr:YSO than Eu:YSO, we observe a large parameter range for both systems over which the EIT width depends only on the control Rabi frequency $\Omega$ and the optical inhomogeneous width $\sigma_{\mathrm{opt}}$. These results quantitatively agree with a theoretical treatment of the system, suggesting that a large class of inhomogeneously broadened systems exhibit behavior that does not depend on the single-atom properties of the individual emitters, but only on the properties of the ensemble as a whole. 


\section{Simple theoretical treatment} \label{theory}
Consider the $\Lambda$-type energy level scheme depicted in Fig. \ref{fig:levels}(a). The transitions from two long-lived ground states, $|1\rangle$ and $|2\rangle$, to a single excited state, $|3\rangle$, are addressed optically with a weak probe field and a strong coupling field, respectively. In many solid-state systems, variations in the local electric field environment cause different emitters to have slightly shifted transition energies. In rare-earth-doped solids, both the optical electronic transition and the ground hyperfine transition are inhomogeneously broadened by this effect. 

We are interested in the transmission of the probe field as a function of detuning for various values of the coupling Rabi frequency $\Omega$ and the inhomogeneous widths of the optical and spin transitions $\sigma_{\mathrm{opt}}$ and $\sigma_{\mathrm{spin}}$, respectively. In a typical homogeneously broadened system (where decay rate, including dephasing, on the $|i\rangle\rightarrow|j\rangle$ transition is denoted $\gamma_{ij}$), we can follow the treatment in \cite{MF05} to obtain the shape of the linear susceptibility $\chi$ of a weak probe field in the perturbative regime:
\begin{equation}
\begin{split}
\chi&\propto\frac{2i\gamma_{21}+4\delta}{\Omega^2+(\gamma_{21}-2i\delta)(\gamma_{31}-2i\Delta)}
\label{Eqn:chi}
\end{split}
\end{equation} 
where parameter definitions can be found in Fig 1(a) and its caption. For a resonant control field ($\delta=\Delta$), this expression for the susceptibility exhibits two important features. First, the imaginary part (which is proportional to the absorption of the probe field) has a transparency window around the two-photon resonance ($\delta=0$) whose width can be smaller than the natural linewidth of the probe transition ($\gamma_{31}$). Second, the real part (which describes the dispersion of the medium) has a sharp slope around the two-photon resonance that leads to significantly reduced probe group velocity. The appearance of a narrow transparency window and slowing of light are the hallmarks of EIT \cite{fleischhauer2000quantum}.

We now consider an inhomogeneously broadened ensemble, which can be thought of as a collection of homogeneously broadened ensembles, each with some shift of its transition energy. We have parameterized the detunings relative to state $\left|1\right\rangle$ such that a shift of the spin transition energy, $\delta_s$, affects the two-photon detuning, $\delta$, but not the probe detuning and the shift of the optical transition energy, $\delta_o$, affects the probe detuning, $\Delta$, but not the two-photon detuning. Thus, $\delta\rightarrow\delta-\delta_s$ and $\Delta\rightarrow\delta-\delta_o$ (where we assume the control field is centered on the optical inhomogeneous line). We have assumed uncorrelated shifts of the optical and spin transitions for each atom in the ensemble, which is the case for the rare-earth solids studied here \cite{ahlefeldt2016ultranarrow}. The susceptibility of the inhomogeneous system (denoted $\tilde{\chi}$) is the homogeneous susceptibility integrated over the inhomogeneous profiles $P_o(\delta_o)$ and $P_s(\delta_s)$. 
\begin{equation}
\begin{split}
\chi&\propto\frac{2i\gamma_{21}+4(\delta-\delta_s)}{\Omega^2+(\gamma_{21}-2i(\delta-\delta_s))(\gamma_{31}-2i(\delta-\delta_o))}\\
\tilde{\chi}(\delta)&\propto\iint P_o(\delta_o)P_s(\delta_s)\chi(\delta,\delta_o,\delta_s) d\delta_od\delta_s\\
\label{Eqn:int}
\end{split}
\end{equation} 
By inspection, we see that the expression for $\tilde{\chi}$ can be integrated analytically if we assume Lorentzian inhomogeneous profiles with full widths at half maximum (FWHM) $\sigma_{\mathrm{opt}}$ and $\sigma_{\mathrm{spin}}$. This results in a familiar expression for the probe susceptibility in the inhomogeneously broadened system:
\begin{equation}
\begin{split}
\tilde{\chi}(\delta)&\propto\frac{2i(\gamma_{21}+\sigma_{\mathrm{spin}})+4\delta}{\Omega^2+((\gamma_{21}+\sigma_{\mathrm{spin}})-2i\delta)((\gamma_{31}+\sigma_{\mathrm{opt}})-2i\delta)}.
\label{Eqn:lor}
\end{split}
\end{equation} 
This is the same expression as for the susceptibility of the homogeneously broadened system with the replacements $\gamma_{21}\rightarrow\gamma_{21}+\sigma_{\mathrm{spin}}$ and $\gamma_{31}\rightarrow\gamma_{31}+\sigma_{\mathrm{opt}}$. In the limit of inhomogeneous linewidths much larger than their homogeneous counterparts, we have simply replaced the homogeneous values with the inhomogeneous values. This means that we can use all of our intuition and understanding of EIT in homogeneously broadened systems, including scaling of the bandwidth, group velocity, and visibility, and the crossover from an EIT-like regime where the transparency window is narrower than the optical linewidth, to an Autler-Townes-like regime where the absorption feature is split into two features separated by more than their widths \cite{PMA11}. We discuss later the effect on the susceptibility of deviation from a Lorentzian inhomogeneous profile.

To compare with experiment, we extract expressions for the FWHM and visibility of the EIT feature, $\Gamma_{\mathrm{EIT}}$ and $V_{\mathrm{EIT}}$. The visibility is defined such that $1-V_{\mathrm{EIT}}$ is the residual probe absorption as a fraction of the absorption without the coupling field. We find that these quantities depend only on $\Omega$, $\sigma_{\mathrm{opt}}$, and $\sigma_{\mathrm{spin}}$ in the limit of large inhomogeneous widths ($\sigma\gg\gamma$). Furthermore, we expand to first order in $\sigma_{\mathrm{spin}}$ for the width, as it is smaller than the other relevant quantities in our systems. 
\begin{equation}
\begin{split}
V_{\mathrm{EIT}}&=\frac{\Omega^2}{\Omega^2+\sigma_{\mathrm{opt}}\sigma_{\mathrm{spin}}}\\
\Gamma_{\mathrm{EIT}}&=\frac{\sqrt{\sigma_{\mathrm{opt}}^2+4\Omega^2}-\sigma_{\mathrm{opt}}}{2}\left(1+\frac{\sigma_{\mathrm{spin}}(\sigma_{\mathrm{opt}}^2-\Omega^2)}{\Omega^2\sqrt{\sigma_{\mathrm{opt}}^2+4\Omega^2}}\right)\\
\Gamma_{\mathrm{EIT}}&\approx\frac{\Omega^2}{\sigma_{\mathrm{opt}}}+\sigma_{\mathrm{spin}}\ \ \ \mathrm{for }\ \Omega\ll\sigma_{\mathrm{opt}}
\\
\Gamma_{\mathrm{EIT}}&\approx\Omega-\frac{\sigma_{\mathrm{opt}}+\sigma_{\mathrm{spin}}}{2}\ \ \ \mathrm{for }\ \Omega\gg\sigma_{\mathrm{opt}}
\end{split}
\label{Eqn:EIT}
\end{equation}

We recover the well-known narrowing of EIT in the presence of inhomogeneous broadening \cite{EK02,scherman2012enhancing}. We further recover two distinct regimes where the width scales as the square of the Rabi frequency ($\Omega\ll\sigma_{\mathrm{opt}}$, EIT regime) and linearly with the Rabi frequency ($\Omega\gg\sigma_{\mathrm{opt}}$, Autler-Townes regime) \cite{PMA11}. Consider a value of $\Omega$ in the Autler-Townes regime for a homogeneously broadened system ($\Omega\gg\gamma_{31}$), but far from such a regime in the inhomogeneously broadened system ($\Omega\ll\sigma_{\mathrm{opt}}$). Rather than two absorption peaks split by $\approx\Omega$, the inhomogeneous system exhibits a transparency window that resembles a homogeneously broadened system in the EIT regime with linewidth $\sigma_{\mathrm{opt}}$ (and is thus narrower than the naively expected width of $\Omega$ by a factor of $\approx\Omega/\sigma_{\mathrm{opt}}\ll1$). Reaching the regime with two well-separated absorption peaks requires $\Omega\gg\sigma_{\mathrm{opt}}$. This limit is difficult to reach in many Doppler-broadened gases, but we see clear Autler-Townes behavior in our rare-earth doped crystals with controllable inhomogeneous broadening (see Fig. \ref{fig:EIT} (b)). Similarly, we see in the visibility the expected behavior that the EIT disappears for $\Omega^2\ll\sigma_{\mathrm{opt}}\sigma_{\mathrm{spin}}$. 

\section{Experimental setup}
We investigate EIT in two different cryogenically cooled rare-earth-doped solids. These are a 0.01\% doped $\mathrm{Eu^{3+}:Y_2SiO_5}$ crystal (Eu:YSO) and a 0.05\% doped $\mathrm{Pr^{3+}:Y_2SiO_5}$ crystal (Pr:YSO), each held at $\approx4$~K in a closed-cycle cryostat and addressed by its own frequency doubled diode laser on the $^7F_0\rightarrow{^5D_0}$ transition at 580~nm for Eu:YSO and the $^3H_4\rightarrow{^1D_2}$ transition at 606~nm for  Pr:YSO (Fig. \ref{fig:levels}(b)). Each diode laser is frequency stabilized to a reference cavity and the laser linewidths are $<4~\mathrm{kHz}$ and $<1~\mathrm{kHz}$ for the Eu:YSO and Pr:YSO transitions, respectively. We note that non-zero laser linewidth has the same effect on EIT as spin inhomogeneity, so $\sigma_{\mathrm{spin}}$ is the quadrature sum of the laser linewidth and the intrinsic spin inhomogeneous width \cite{BL97}. For each rare-earth-doped crystal, the probe and coupling fields are derived from the same laser and given a relative frequency shift with acousto-optic modulators in a double-pass configuration. The probe and coupling fields intersect in the crystal at an angle of $<$2$^{\circ}$ that allows the $\approx0.5$ mm diameter fields to overlap for the entire 10~mm length of the crystal. The probe field frequency is scanned over the two-photon resonance and the transmitted intensity is recorded as a function of time, which is then converted to frequency. The scan speed is kept below a rate equal to the spin inhomogeneous width divided by the optical coherence time to avoid coherent oscillations \cite{FM12}.

The two systems differ primarily in the strength of the optical transition, which is substantially weaker and longer-lived in Eu:YSO compared to Pr:YSO. As a result, the parameter regimes we can access allow extremely narrow EIT in Eu:YSO and clear Autler-Townes-type behavior in Pr:YSO. Furthermore, coherent transients in the EIT are more apparent in Eu:YSO, which may be useful for EIT based-sensing or other applications \cite{FM12}.

The Rabi frequency of the control field is calibrated from the frequency of the observed optical nutation when the medium is suddenly illuminated by the control field \cite{YS00}. The optical inhomogeneous width is determined from the absorption of the weak probe as a function of detuning in the absence of the coupling field (see below for more details on controlling the optical inhomogeneity). The spin inhomogeneous width is an uncontrolled, fixed value that is due to disorder in the crystal. 

\section{Spectral hole burning}
To study the role of optical inhomogeneous broadening in EIT we work in a regime where $\sigma_{\mathrm{opt}}$ and $\Omega$ are similar in value and larger than the uncontrolled $\sigma_{\mathrm{spin}}$ because the EIT visibility drops precipitously for $\Omega^2<\sigma_{\mathrm{opt}}\sigma_{\mathrm{spin}}$ \cite{EK02}. The optical inhomogeneous linewidth of the entire ensemble is $\approx10^7$ times larger than the homogeneous linewidth of the atoms. This broadening is also larger than the hyperfine splittings, which means that some selection of a subensemble of atoms is required to observe coherent effects involving the spin states. We use spectral hole burning techniques to select a subensemble of atoms within a much narrower spectral region and pump atoms nearby in the inhomogeneous profile to the non-participating ground hyperfine level \cite{GP00,MN04}. This technique allows inhomogeneous widths larger than the laser linewidth (integrated over the $\approx\mathrm{ms}$ optical pumping time) and smaller than the hyperfine splittings. In practice, we generate subensembles with spectral widths $10^2$ to $10^4$ times the homogeneous linewidth of the atoms.

\begin{figure}[ht]
  \centering
  \includegraphics[width=\columnwidth]{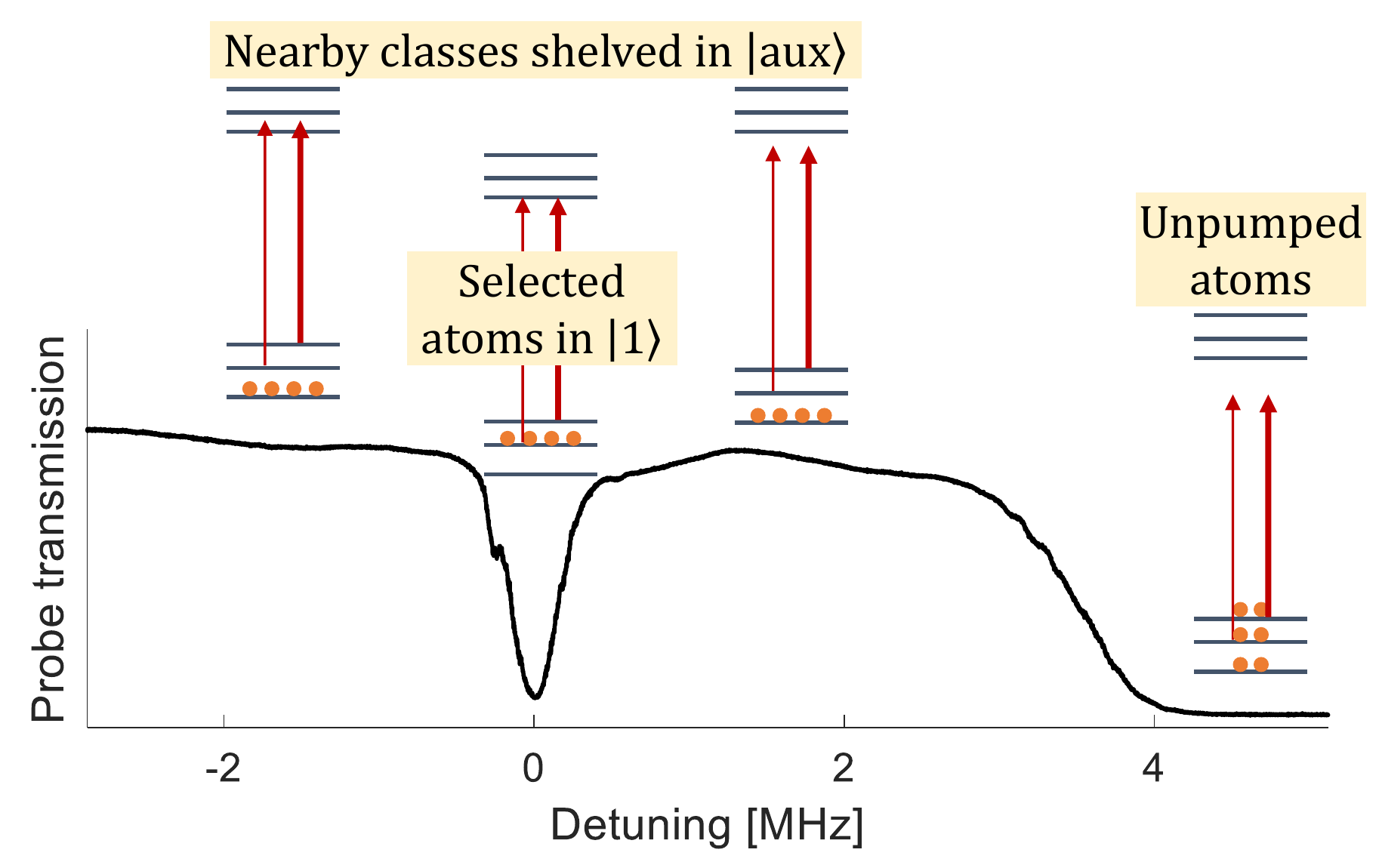}
  \caption{Transmission of a weak probe (with no coupling field) as a function of frequency near zero detuning following state preparation. Energy level structures for chosen frequency class plus three other classes shown with probe (thin arrow) and coupling (thick arrow) fields for reference.}  
  \label{fig:hole_burning}
\end{figure}

We use a multi-step spectral hole-burning sequence to prepare a sub-ensemble of atoms in a particular frequency class and ground hyperfine state, while pumping atoms in other nearby frequency classes to ground states such that they are far from resonance with the probe and control fields \cite{GP00,MN04}. An example hole-burned absorption profile is shown in Fig. \ref{fig:hole_burning}.

Both Pr:YSO and Eu:YSO have electron spin-singlet ground and excited states that are split into three doubly degenerate nuclear hyperfine states in zero magnetic field with splittings in the range of $\approx$5~MHz to 100~MHz. Praseodymium has a single naturally occurring isotope, while europium has two isotopes that occur naturally in approximately the same abundance and have different hyperfine structures. All ground to excited transitions are allowed with varying transition strengths in both systems and all fields are linearly polarized along the crystallographic axis that maximizes the light-matter interaction \cite{GL06}. The existence of a third metastable ground state is important as it acts as an auxiliary state where unwanted population can be shelved to allow coherent processes on the other two states. In both Eu:YSO and Pr:YSO, we use the upper two ground states for EIT and the lowest as the auxiliary state.

The large inhomogeneous broadening of the full ensemble means that at any optical frequency within the inhomogeneous bandwidth there are atoms in nine different frequency classes resonant on each of the nine different transitions. (In the natural abundance europium used here, there are an additional nine frequency classes of the other isotope resonant on its nine transitions). The first step in the hole burning procedure is selecting a single frequency class of interest by applying three fields at frequencies such that the chosen frequency class is resonant with all three fields on transitions from each of the ground states. All other frequency classes can be resonant with at most two of the fields and will be optically pumped out of the ground state(s) with a resonance that matches a resonance of the chosen frequency class. We sweep these fields over a range much larger than the ultimate desired subensemble to prepare a spectral region with increased transparency. We then empty out the two ground states that make up the $\Lambda$ system by turning off the third field that is at neither the probe nor control frequency. Finally, we repopulate a narrow spectral region in $|1\rangle$ with a single frequency repump field, while keeping the field at the control frequency on to prevent population build up in $|2\rangle$. An example trace of the transmission of a weak probe measuring the final absorption profile is shown in Fig. \ref{fig:hole_burning}.

We generate the absorbing feature by illuminating the sample with a repump field for a variable amount of time. The width of the feature depends on the power broadening of the repump light, the spectral diffusion of the atoms during the repump, and the laser linewidth averaged over the repump time. In practice, the laser noise during the $>$100~ms repump sets the width of the absorbing feature, and by controlling the repump time we can vary the final width from $\lesssim300~\mathrm{kHz}$ to $\gtrsim2~\mathrm{MHz}$ in both systems. We do not generate wider absorbing features to ensure that the probe transmission is dominated by atoms in the absorbing feature rather than atoms outside the transparency window burned around the probe frequency. In Eu:YSO, the transparency window is limited to $\approx5~\mathrm{MHz}$ by the level structure and in Pr:YSO we create a single transparency window that covers both the probe and control frequencies, which leaves the absorbing feature $\approx3~\mathrm{MHz}$ from the edge of the window.  

\section{Experimental results}
We measure the EIT over a range of both optical inhomogeneous widths ($\sigma_{\mathrm{opt}}$) and control Rabi frequencies ($\Omega$) in both the Eu:YSO and Pr:YSO systems. With the control field centered on the hole-burned feature, the probe field is scanned in frequency across the two-photon resonance. The probe transmission is recorded as a function of time and converted to transmission as a function of frequency. We note that coherent effects on the probe transmission limit the speed of the frequency scan, particularly in Eu:YSO, where the optical transition has a coherence time $>1~\mathrm{ms}$.

\begin{figure}[ht]
  \centering
  \includegraphics[width=\columnwidth]{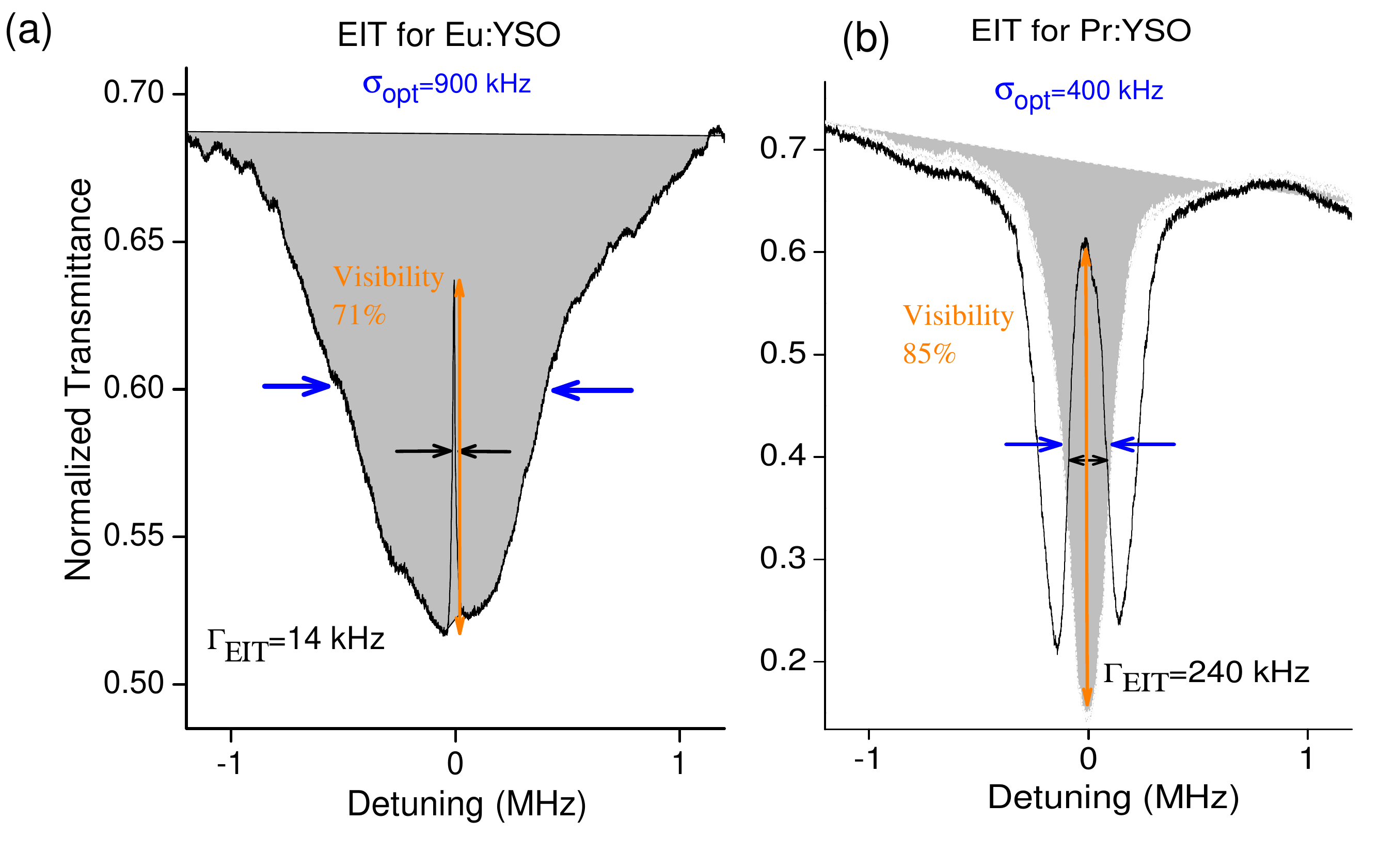}
  \caption{Example measured EIT spectra (black lines) in (a) Eu:YSO and (b) Pr:YSO. The shaded area is the hole-burned feature. The optical inhomogeneous width and measured EIT linewidth are noted on each figure.} 
  \label{fig:EIT}
\end{figure}

The available laser power for each system allows us to study EIT for Eu:YSO only for $\Omega\ll\sigma_{\mathrm{opt}}$, while in Pr:YSO we can reach $\Omega\gtrsim\sigma_{\mathrm{opt}}$. For $\Omega\ll\sigma_{\mathrm{opt}}$, the shape of the EIT transmission peak is approximately Lorentzian and its width can be extracted by fitting. Outside this regime, the overall absorption appears as two separated peaks and we extract the FWHM without fitting any particular shape to the feature. The optical depth of the Eu:YSO system is sufficiently small ($<2$ for all prepared features) that the visibility can be directly extracted from the probe transmission as the ratio of the transmission at the EIT peak to the transmission away from the hole-burned feature. The larger optical depth in the Pr:YSO ($\gtrsim6$ for the widest hole-burned features) necessitates fitting a saturated absorption function to the hole-burned features in order to extract the visibility. Figure~\ref{fig:EIT} shows typical EIT features in two different parameter regimes. For $\Omega\ll\sigma_{\mathrm{opt}}$ the EIT window is at the center of the hole-burned absorbing feature, while for $\Omega\sim\sigma_{\mathrm{opt}}$ the absorbing feature appears split by the control field as typical of Autler-Townes splitting. We note that at the largest EIT widths we observe a small absorption peak at zero detuning (not shown). This can be attributed to atoms in frequency classes outside the spectral hole burned region. We confirm this effect by performing the integration in Eq. \ref{Eqn:int} numerically, with $P_o(\delta_o)$ that includes an absorbing feature at the center of a hole-burned trench with a broad inhomogeneous ensemble outside. 

\begin{figure}[ht]
  \centering
  \includegraphics[width=\columnwidth]{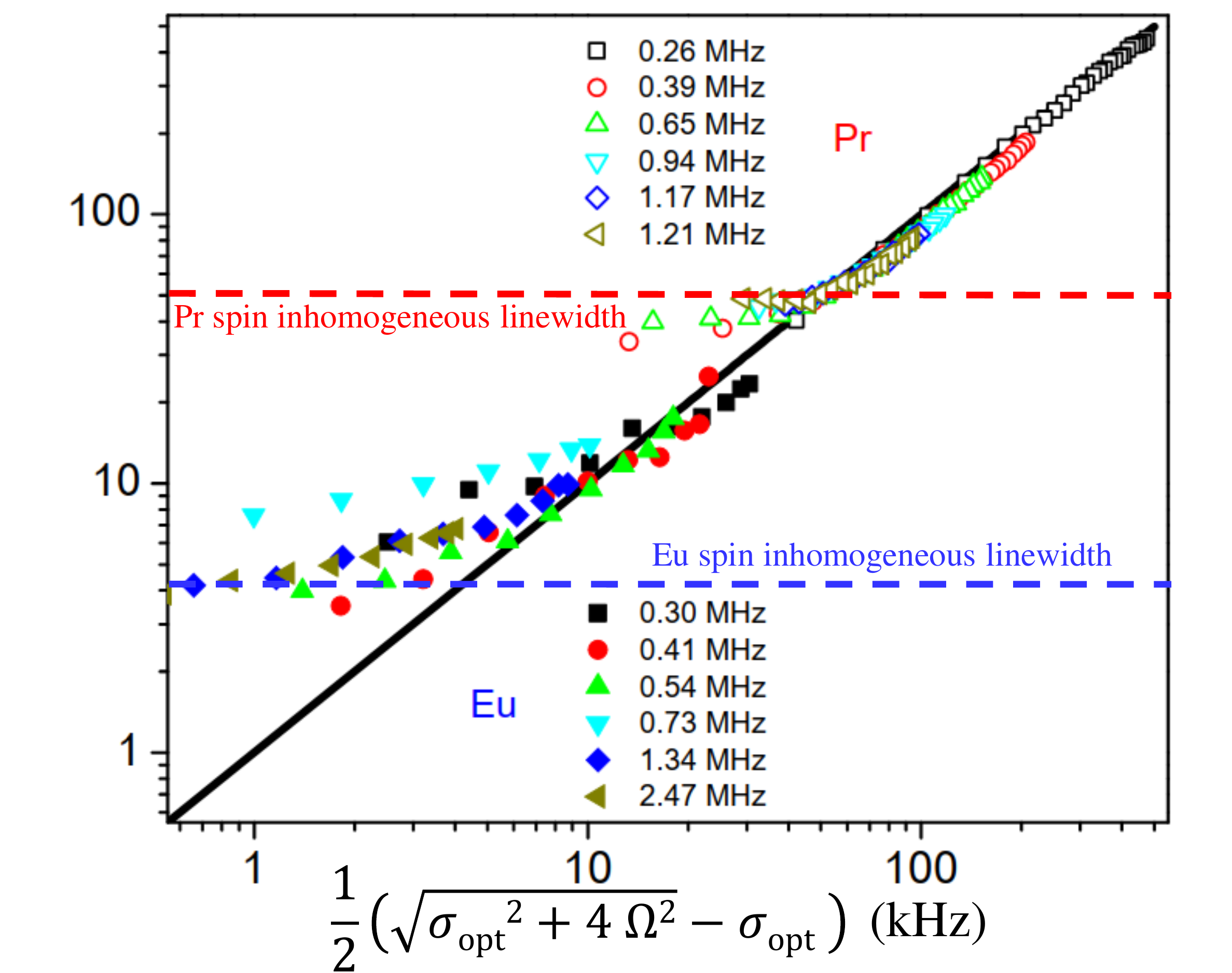}
  \caption{ Measured vs. theoretical EIT width for Eu:YSO (solid markers) and Pr:YSO (hollow markers) at different Rabi frequencies and optical inhomogeneous widths. Optical inhomogeneous widths are as indicated. Solid line is the unit slope. The spin inhomogeneity in each system (horizontal dashed lines).} 
  \label{fig:combined}
\end{figure}

The measured width of the transparency window is plotted against the expected value $\frac{1}{2}(\sqrt{\sigma_{\mathrm{opt}}^2+4\Omega^2}-\sigma_{\mathrm{opt}})$ in Fig. \ref{fig:combined}. We have ignored the term in the expected value that depends on the spin inhomogeneity (Eq. \ref{Eqn:EIT}), which matters only for the smallest Rabi frequencies studied and accounts for the deviation of the data from the unit slope line at small values. The value of the spin inhomogeneity for each system is noted in Fig. \ref{fig:combined} with dashed horizontal lines. The Eu:YSO value of $4~\mathrm{kHz}$ is similar to the measured laser linewidth, suggesting that the intrinsic spin inhomogeneity is smaller than a previously measured value in a similar sample \cite{NT13}. The Pr:YSO value of $40~\mathrm{kHz}$ is inferred to be the intrinsic spin inhomogeneous width, as it is much larger than the measured laser linewidth and consistent with previously measured values \cite{BSH97a}. 

\begin{figure}[t]
  \centering
  \includegraphics[width=\columnwidth]{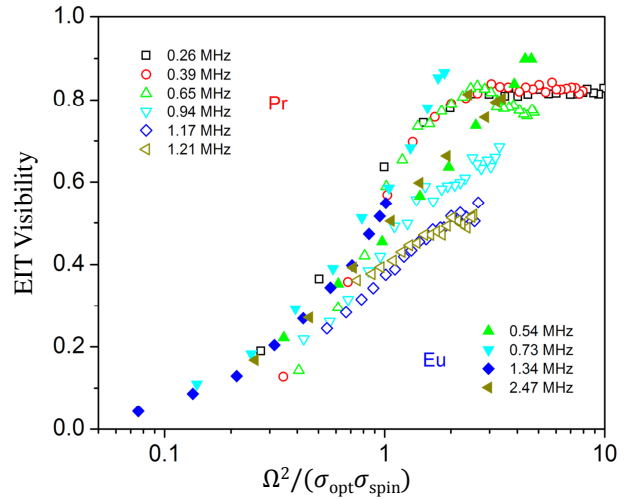}
  \caption{ Measured EIT width for Eu:YSO (solid markers) and Pr:YSO (hollow markers) at different optical inhomogeneous widths. Optical inhomogeneous widths are as indicated.} 
  \label{fig:visibility}
\end{figure}

The visibility for several different configurations and both species similarly collapse as a function of $\Omega^2/(\sigma_{\mathrm{opt}}\sigma_{\mathrm{spin}})$, as seen in Fig. \ref{fig:visibility}. We see the expected behavior that the visibility rises from zero to near one, crossing 0.5 near the expected value of $\Omega^2\approx(\sigma_{\mathrm{opt}}\sigma_{\mathrm{spin}})$ for both species. The saturation of the visibility at large Rabi frequency is likely due to residual absorption of far off resonant atoms. This collapse offers strong support for our model because of the order of magnitude difference in the value of $\sigma_{\mathrm{spin}}$ for europium and praseodymium. 

We observe that the data from the two different systems follows the same scaling law that depends only on the control field Rabi frequency and optical inhomogeneous width. The single atom properties of each system, namely the homogeneous linewidth, does not affect the EIT linewidth. Thus, properties like the optical lifetime and coherence time are independent of the bulk ensemble response to the probe and coupling fields. Both the EIT width and visibility exhibit same scaling with $\Omega$, $\sigma_{\mathrm{opt}}$, and $\sigma_{\mathrm{spin}}$ for the two different rare-earth species studied over two orders of magnitude. This occurs despite the different values of the single atom (homogeneous) parameters for each system.


\section{Non-Lorentzian inhomogeneous profiles}
Most real systems do not exhibit the Lorentzian inhomogeneous profile we assumed in section \ref{theory}. In Doppler broadened gases the optical inhomogeneity is Gaussian, and the spectral hole-burned features here have a range of shapes depending on the specific implementation. In this work, we extract the FWHM of the hole-burned spectral features without assuming any particular shape. Thus, it is important to consider the validity of our theoretical treatment for non-Lorentzian lineshapes.

The integral in Eq. \ref{Eqn:int} cannot in general be calculated analytically for inhomogeneous profiles with non-Lorentzian distributions. In order to gain an understanding of the role of the distribution shape, we perform numeric integration over Gaussian and flat-top profiles for a range of parameters to obtain $\tilde{\chi}(\delta)$. We then extract the FWHM of the EIT feature and the EIT visibility (defined as the difference of the maximum and minimum values of the imaginary part of $\tilde{\chi}$ divided by their sum). 

\begin{figure}[htb]
  \centering
  \includegraphics[width=\columnwidth]{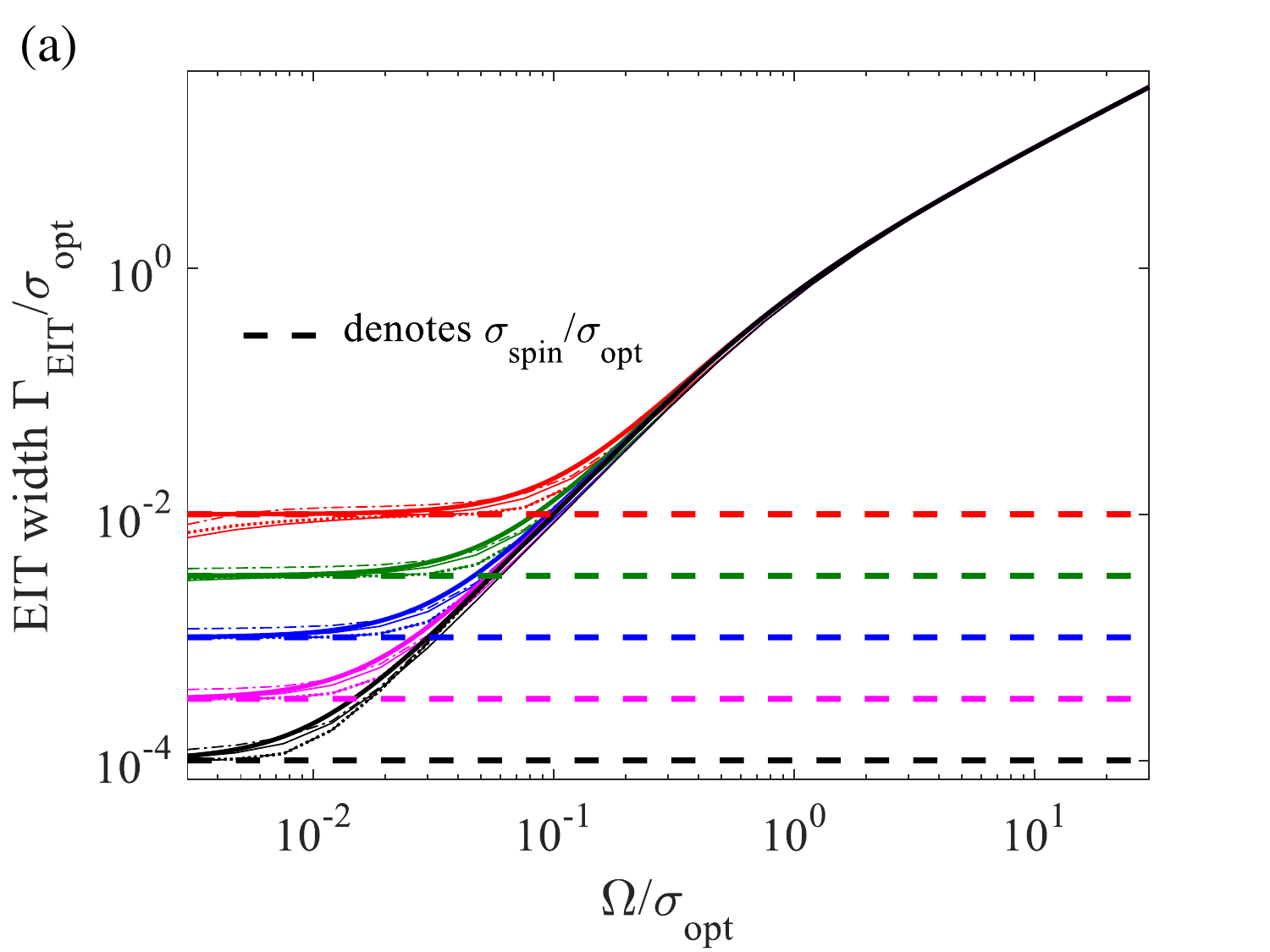}
  \includegraphics[width=\columnwidth]{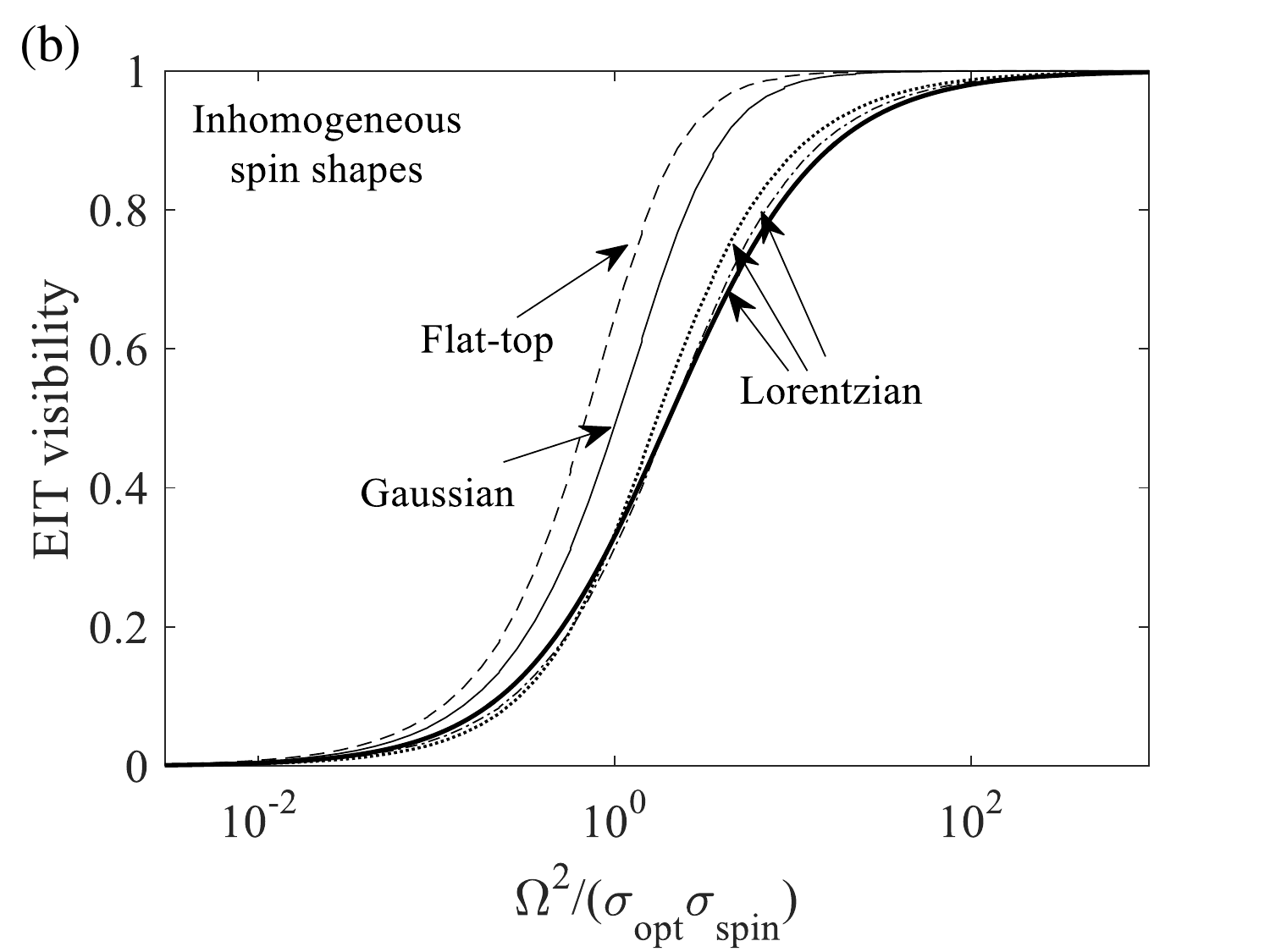}
    \caption{(a) EIT width vs $\Omega/\sigma_{\mathrm{opt}}$ and (b)~EIT visibility vs $\Omega^2/(\sigma_{\mathrm{opt}}\sigma_{\mathrm{spin}})$ are both extracted from Eq. \ref{Eqn:EIT} and numerical integration over Gaussian and flat-topped inhomogeneous profiles. In (a) different colors correspond to different values of the spin inhomogeneity, $\sigma_{\mathrm{spin}}/\sigma_{\mathrm{opt}}$, (the values of which are denoted with horizontal dashed lines) and different curves correspond to different inhomogeneous shapes (analytical results for Lorentzian shapes are thick solid lines). Different curves are nearly indistinguishable and correspond to different spin and optical inhomogeneous shapes showing the insensitivity to those shapes. In (b), the leftmost results are integrated over flat-topped (dashed line) and Gaussian (thin solid line) spin inhomogeneous shapes. The nearly indistinguishable set of curves to the right are all Lorentzian spin inhomogeneous broadening with different optical inhomogeneous shapes.}
  \label{fig:sims}
\end{figure}

The results of these numerical integrations are shown in Fig. \ref{fig:sims} along with the analytical results for different values of spin inhomogeneity and inhomogeneity shapes. For the width (see fig. \ref{fig:sims}a), we see the linear dependence at large $\Omega$, quadratic dependence at small $\Omega$ and saturation at the value of the spin inhomogeneity at very small $\Omega$. Similarly for the visibility of the EIT transmission peak (see fig. \ref{fig:sims}b), we see a transition from low transmission when width is saturated, to high transmission when the spin inhomogeneity is negligible compared to the other quantities. Neither the linewidth of the EIT nor the visibility depends on the shape of the optical inhomogeneity. The only dependence suggested by the numerical results is increased EIT transmission at smaller $\Omega$ for spin inhomogeneous broadening that falls off more quickly than Lorentzian. Thus, the replacement of the homogeneous linewidth with its inhomogeneous counterpart is thus a reasonable technique for considering EIT in a wide range of inhomogeneous broadened ensembles.

\section{Conclusion}
In conclusion, we have studied EIT in two different inhomogeneously broadened rare-earth doped solids. As opposed to ensembles of identical or near-identical cold atoms (homogeneously broadened ensembles), most solids exhibit large inhomogeneous broadening. As solid-state systems become more common for quantum optics and quantum information applications due to their advantages in terms of of motional dephasing, reduced experimental overhead, and integratability into scalable photonic systems, it is vital to explore and understand the impact of inhomogeneous broadening. Here we observe good agreement with a theoretical treatment covering two orders of magnitude in the coupling Rabi frequency and inhomogeneous linewidth. In addition, a simple theoretical treatment of inhomogeneous broadening predicts EIT lineshapes similar to those seen in homogeneously broadened systems, with a direct replacement of the homogeneous linewidths with their inhomogeneous counterparts. We further discuss the effect of the shape of the inhomogeneous profile on the EIT properties, and see that the properties of interest are largely insensitive to the shape. This work provides important groundwork for implementing coherent quantum optical processes in solids where inhomogeneous broadening often plays a major role, thus paving the way to exploiting this class of materials for quantum information applications. 

\section*{Acknowledgments}
We acknowledge partial support from the NSF Physics
Frontier Center at the Joint Quantum Institute. The authors thank Paul Lett and Ivan Burenkov for helpful discussions. 

\end{document}